\documentclass[12pt,thmsa]{article}
\def \eq {\begin{equation}}
\def \fim-eq {\end{equation}}
\begin{document}

\author{E. S. Guerra \\
%EndAName
Departamento de F\'{\i}sica \\
Universidade Federal Rural do Rio de Janeiro \\
Cx. Postal 23851, 23890-000 Serop\'edica, RJ, Brazil \\
email: emerson@ufrrj.br\\
}
\title{REALIZATION OF GHZ STATES AND THE GHZ TEST VIA CAVITY QED}
\maketitle

\begin{abstract}
\noindent In this article we discuss the realization of atomic GHZ states
involving three-level atoms and we show explicitly how to use this state to
perform the GHZ test in which it is possible to decide between local realism
theories and quantum mechanics. The experimental realizations proposed makes
use the interaction of Rydberg atoms with a cavity prepared in a coherent
state.

\ \newline

\end{abstract}

%\date{The Date }

\section{INTRODUCTION}

Quantum mechanics has given rise to many questions of philosophical nature,
for instance, about the concept of reality \cite{JammerPhil, Omnes, Hughes,
Krips, Espagnat}. "Objective reality" (the reality of objects outside
ourselves) seems, as Heisemberg put, to have \ "evaporated" as a result of
quantum physics. Based on the intuitive notion that the world outside the
self is real and has at least some properties that exists independent of the
human consciousness we would expect that \ "objective reality" would
maintain. The counterintuitive concepts and notions in quantum mechanics are
not easy to accept without any questioning and spent of time thinking about
so strange behavior which nature reveals to us theoretically and
experimentally, which gives us a completely new view of the reality. Even
nowadays there are many interpretations of the quantum theory \cite%
{JammerPhil, Omnes, Hughes, Krips} and there is no consensus among which\ of
the several interpretations is the correct one and there are still some
points to be clarified. This conterintuitive concepts and their consequences
lead some outstanding scientists as, for instance Einstein, to question the
theory.

Entanglement is a feature of quantum mechanics which is fascinating but, on
the other hand, yields as a consequence non-locality which contradicts
completely our commom sense. Einstein, Podolsky and Rosen, noticed the
implications of quantum entanglement and proposed a \textit{gedanken }%
experiment, the EPR experiment \cite{EPR, Omnes, Hughes, Krips, WallsMilburn}
in a fine article which deserves some comments. Although the EPR experiment
was imagined to show that quantum theory was not a complete theory to
describe reality ironically it turned out to be a very important
contribution to the theory. The EPR strategy was to describe an experimental
arrangement involving correlated pair of particles (EPR states or Bell
states). These particles interact and then are separated. Therefore,
measurements made on one particle can be used via correlation to generate
predictions about the other particle. In the EPR thought experiment the
incompatible observables in question are position and momentum. In their
article EPR state that the physical concepts with which a theory operates
are intended to correspond with the objective reality. \ EPR look at the
mathematical model supplied by quantum mechanics and give us a sufficient
condition for an element of that model to represent an element of reality.
They state: \ "If, without in any way disturbing a system we can predict
with certainty the value of a physical quantity, then there exists an
element of physical reality corresponding to this physical quantity". As a
necessary condition for the completeness of the theory, EPR state that \
"every element of the physical reality must have a counter part in the
physical theory". According to the EPR claim, for instance, if the
non-commuting spin-$\frac{1}{2}$ observables $S_{z}$ and $S_{x}$ had both
simultaneous reality, and thus definite values, these values would enter
into the complete description, according to the condition of completeness.
If then the wave function provided a complete description of reality, it
would contain these values. However, according to quantum mechanics the spin
state vector cannot "contain" the values of $S_{z}$ and $S_{x}$
simultaneously. For instance, in the eigenstate $|z,+\rangle $ ($z$
component of the spin up) of $S_{z}$ the particle has the property $%
(S_{z},+) $ and the value of $S_{z}$ is predictable with certainty, and so
there is an element of reality corresponding to it. However, we could also
say that, in this state, the particle has neither the property \ $(S_{x},+)$
\ nor the property $(S_{x},-)$ \ since the particle state could be projected
in the eigenstates $|x,+\rangle $ and $|x,-\rangle $ with the same
probability $1/2$. That is, neither of these properties constitute an
element of reality. EPR saw that the fact that quantum mechanics admits no
dispersion-free states does not, on its own, tell us wether the theory is
complete or not. In their words: \ "From [the dispersion principle] it
follows (1) the quantum mechanics description of reality given by the wave
function is not complete or (2) when operators corresponding to two physical
quantities do not commute the two quantities cannot have simultaneous
reality". Shortly after the EPR paper Bohr published a reply in which he
defends the completeness of the quantum-mechanical description of nature
which could not be refuted properly by EPR\ and which came to be known as
the Copenhagen interpretation \cite{BohrEPRreply}. A modified version of the
EPR experiment, however analogous, was suggested by Bohm \cite{BohmEPR,
Omnes, Hughes}. In Bohm's experiment the observables are different
components of spin of the spin-$\frac{1}{2}$ particle. For instance, the EPR
pair could be an electron-positron pair or twin photons emitted by excited
atoms in a two-photon transition in which the photons are emitted with
orthogonal polarizations. There are nowadays several experimental proposals
of such experiment and some have been performed in laboratory (see for
instance \cite{Zeilinger}). A proposal of generating EPR states via cavity
QED is presented, for instance, in \cite{GHZWalther, RevHaroche}. Hidden\
variable theories \cite{Omnes, Hughes} were \ proposed in which one would
recover local realism and one of its proponents was Bohm \ \cite{Bohmhvt,
Omnes, Hughes}. \ Bell developed a clever theoretical tool which could be
used to perform experiments based on it which could decide between local
realism theories and quantum mechanics, the Bell's inequality \cite{Bell,
Omnes, Hughes, WallsMilburn}. In the 1980s it was realized experiments by
Aspect and collaborators \cite{Aspect, WallsMilburn} based on the Bell's
inequality which although strongly favored quantum mechanics, there remained
some possibility \ for which a local reality view could still be maintained.
Concerning the experimental works developed to test Bell's inequality, \ in
1981 Bell stated: \ "It is hard for me to believe that quantum mechanics
works so nicely for inefficient practical set-ups, and is yet going to fail
when sufficient refinements are made". We should point out that the main
difficulty in the test of the Bell's inequality is related to the fact that
there is no perfect particle detectors. For a review of Bell's inequalities
and some of its variants see \cite{RevBellsIneq}.

As we have mentioned Bell's inequality has been a clever tool in order to
test quantum mechanics. However, in 1989, Greenberger, Horne and Zeilinger
found out an ingenious theoretical tool to test quantum mechanics confronted
with local hidden variable theories, the Bell theorem without inequalities 
\cite{GHZ, Mermin, RevBellsIneq}, which can demonstrate the spookiness of
quantum mechanics even more dramatically than the Bell's analysis. The GHZ
scheme is based on a clever choice of three commuting operators which are
each formed by the product of three particle spin-$\frac{1}{2}$ operators,
two for the $y$ component and one for the $x$ component, and a fourth
operator formed by the product of three particle spin-$\frac{1}{2}$
operators all of them for the $x$ component. The GHZ \ test is performed
applying these operators to a three particle entangled state, the GHZ state.
Contrary to experimental tests involving the Bell's inequalities, the GHZ
test is to be performed in just one run of the experiment. The decision
between quantum mechanics and local hidden variable theories is taken
observing the eigenvalue of the GHZ state for the operator involving the
three particle spin-$\frac{1}{2}$ operators all of them for the $x$
component. Such eigenvalue can be $\pm 1$ since there are two possibilities
for defining the GHZ state. One involving a plus sign and other involving a
minus sign. That is, these states are constructed with the product of the
three spin up states for the three particles $\pm $ the product of the three
spin down states for the three particles. Depending on which GHZ state we
are using the eigenvalue will be $+1$ or $-1$. If the sign in the GHZ state
is $+1$, according to quantum mechanics \ the eigenvalue of the GHZ state \
for the operator involving the three particle operators all of them for the $%
x$ spin component will be $+1,$ and if the sign in the GHZ state is $-1$,
according to quantum mechanics \ this eigenvalue of the GHZ state will be $%
-1 $. However, for instance, for the first case (eigenvalue $+1$) if we
assume that there are elements of reality to be reveled according to a local
hidden variable theory the eigenvalue should be $-1$. As we see, the
decision between quantum mechanics and local realism theories is a binary
simple one showing the power of the GHZ formalism. \ There are several
proposals of preparation of GHZ states. For instance, proposals involving
cavity QED are presented in \cite{GHZLambdaat, GHZWalther, Gerry,
RevHaroche, Zeng, YeGuo1}. A scheme to prepare a two mode cavity EPR state
and a three cavity mode GHZ state is presented in \cite{SongLuZhangGuo}. A
non-maximally entangled GHZ state is used in a scheme of teleportation in 
\cite{DaiChenLi}. A scheme to produce GHZ atomic states and for
teleportation of atomic states is discussed in \cite{YeGuo2}. \ The
realization of a spin-type GHZ state via an atomic interference method is
discussed in \cite{SongGuo}. An interesting proposition of generating EPR
states and realization of teleportation using a dispersive atom-field
interaction where two atoms interact simultaneously with a cavity is
discussed in \cite{ZengGuo}. A scheme of generating GHZ states also using a
dispersive atom-field interaction where three atoms interact simultaneously
with a cavity is presented in \cite{Zeng2}. A method for preparation of an
entangled field state is proposed in \cite{IkramZhuZubairy}. For a concrete
experimental realization of GHZ states see \cite{EXPGHZ, RevHaroche}. As a
last remark about GHZ sates and \ their many particle generalizations we
mention that they are essential ingredients in many quantum communication
and quantum computation schemes \cite{Zeilinger, EXPGHZ, MPGHZ}.

In reference \cite{GHZLambdaat} we have proposed a preparation of atomic GHZ
states and a GHZ test similar the one discussed in this article. However, in
reference \cite{GHZLambdaat} we have used three-level lambda atoms in which
the lower states are degenerate states. The advantage of using lambda atoms
is that it is not necessary to perform many atomic state rotations as in the
case of the non-degenerate system used here. However, the atomic state
rotation and atomic state detection are simpler in the non-degenerate case
used in the present article. This is the main reason we developed the study
reported in this article.

In the sections which follow we assume that the atoms we are going to use
are Rydberg atoms of relatively long radiative lifetimes \cite{Rydat}. We
also assume perfect \ microwave cavities, that is, we neglect effects due to
decoherence. Concerning this point, it is worth to mention that nowadays it
is possible to build up niobium superconducting cavities with high quality
factors $Q$. It is possible to construct cavities with quality factors $%
Q\sim 10^{8}$ \cite{haroche}. Even cavities with quality factors as high as $%
Q\sim 10^{12}$ have been reported \cite{walther}, which, for frequencies $%
\nu \sim 50$ GHZ gives us a cavity field lifetime of the order of a few
seconds. Maybe future technological achievements will allow us to build up
cavities with very high quality factors in which the fields can be stored
for a very long time. This perhaps is one of the main goal of the cavity QED
experimental physicists.

\section{THE EPR EXPERIMENT\protect\bigskip}

Let us consider a three-level cascade atom \ $Ak$ with $\mid e_{k}\rangle
,\mid f_{k}\rangle $ and $\mid g_{k}\rangle $ being the upper, intermediate
and lower atomic states (see Fig. 1). We assume that the transition $\mid
f_{k}\rangle \rightleftharpoons \mid e_{k}\rangle $ is far enough from
resonance with the cavity central frequency such that only virtual
transitions occur between these states. In addition we assume that the
transition $\mid f_{k}\rangle \rightleftharpoons \mid g_{k}\rangle $ is
highly detuned from the cavity frequency so that there will be no coupling
with the cavity field. Here we are going to consider the effect of the
atom-field interaction taking into account only levels $\mid f_{k}\rangle $
and $\mid g_{k}\rangle .$ We do not consider level $\mid e_{k}\rangle $
since it will not play any role in our scheme. Therefore, we have
effectively a two-level system involving states $\mid f_{k}\rangle $ and $%
|g_{k}\rangle $. Considering levels $\mid f_{k}\rangle $ and $\mid
g_{k}\rangle $ \ we can write an effective time evolution operator (see
Appendix) 
\begin{equation}
U_{k}(t)=e^{i\varphi a^{\dagger }a}\mid f_{k}\rangle \langle f_{k}\mid
+|g_{k}\rangle \langle g_{k}\mid ,  \label{U}
\end{equation}%
where the second term above was put by hand just in order to take into
account the effect of level $\mid g_{k}\rangle $.

Now, consider first an atom $A1$ is prepared in the Ramsey cavity $R1$ in a
superposition. If atom $A1$ is initially in the state $\mid g_{1}\rangle $,
according to the rotation matrix%
\begin{equation}
R=\frac{1}{\sqrt{2}}\left[ 
\begin{array}{cc}
1 & 1 \\ 
-1 & 1%
\end{array}%
\right] ,  \label{R}
\end{equation}%
we have 
\begin{equation}
\mid \psi \rangle _{A1}=\frac{1}{\sqrt{2}}(\mid f_{1}\rangle +\mid
g_{1}\rangle ).
\end{equation}%
Now, let us assume that we have a cavity $C$ prepared in coherent state $%
|\alpha \rangle $. A coherent state $|\beta \rangle $ is obtained applying
the displacement operator $D(\beta )=e^{(\beta a^{\dag }-\beta ^{\ast }a)}$
to the vacuum, that is, $|\beta \rangle =D(\beta )|0\rangle ,$ and is given
by 
\begin{equation}
|\beta \rangle =e^{-\frac{1}{2}|\beta |^{2}}{\sum\limits_{n=0}^{\infty }}%
\frac{(\beta )^{n}}{\sqrt{n!}}|n\rangle
\end{equation}%
\cite{Louisell, Orszag, WallsMilburn}. Experimentally, it is obtained with a
classical oscillating current in an antenna coupled to the cavity. Let us
take $\varphi =\pi $. Then, \ according to (\ref{U}), the system $A1-C$
evolves to%
\begin{equation}
\mid \psi \rangle _{A1-C}=\frac{1}{\sqrt{2}}(\mid f_{1}\rangle |-\alpha
\rangle +\mid g_{1}\rangle |\alpha \rangle ),
\end{equation}%
where we have used $e^{za^{\dagger }a}|\alpha \rangle =|e^{z}\alpha \rangle $
\cite{Louisell}$.$ Now, if atom $A1$ enters a second Ramsey cavity $R2$
where the atomic states are rotated according to the rotation matrix (\ref{R}%
), we have 
\begin{eqnarray}
&\mid &f_{1}\rangle \rightarrow \frac{1}{\sqrt{2}}(\mid f_{1}\rangle -\mid
g_{1}\rangle ),  \nonumber \\
&\mid &g_{1}\rangle \rightarrow \frac{1}{\sqrt{2}}(\mid f_{1}\rangle +\mid
g_{1}\rangle ),
\end{eqnarray}%
and, therefore,%
\begin{equation}
\mid \psi \rangle _{A1-C}=\frac{1}{2}[\mid f_{1}\rangle (|\alpha \rangle
+|-\alpha \rangle )+\mid g_{1}\rangle (|\alpha \rangle -|-\alpha \rangle )].
\end{equation}%
It is worth to mention at this point that if we define the non-normalized
even and odd coherent states 
\begin{eqnarray}
|+\rangle &=&|\alpha \rangle +|-\alpha \rangle ,  \nonumber \\
|-\rangle &=&|\alpha \rangle -|-\alpha \rangle ,  \label{EOCS}
\end{eqnarray}%
with $N^{\pm }=\langle \pm \mid \pm \rangle =2\left( 1\pm e^{-2\mid \alpha
\mid ^{2}}\right) $ \ and $\langle +\mid -\rangle =0$ \cite{EvenOddCS}, we
have already an EPR state involving the atomic states of $A1$ and the cavity
field state, that is we have%
\begin{equation}
\mid \psi \rangle _{A1-C}=\frac{1}{2}[\mid f_{1}\rangle |+\rangle +\mid
g_{1}\rangle |-\rangle ].
\end{equation}%
Now, let us prepare a two-level atom $A2$ in the Ramsey cavity $R3$. If atom 
$A2$ is initially in the state $\mid g_{2}\rangle $, according to the
rotation matrix (\ref{R}), we have%
\begin{equation}
\mid \psi \rangle _{A2}=\frac{1}{\sqrt{2}}(\mid f_{2}\rangle +\mid
g_{2}\rangle ),
\end{equation}%
and let us send this atom through cavity $C$. Taking into account (\ref{U})
with $\varphi =\pi $, after the atom has passed through the cavity we get%
\begin{equation}
\mid \psi \rangle _{A1-A2-C}=\frac{1}{2\sqrt{2}}[\mid f_{1}\rangle (\mid
f_{2}\rangle +\mid g_{2}\rangle )(|\alpha \rangle +|-\alpha \rangle )-\mid
g_{1}\rangle (\mid f_{2}\rangle -\mid g_{2}\rangle )(|\alpha \rangle
-|-\alpha \rangle )].
\end{equation}%
Then, atom $A2$ enters a Ramsey cavity $R4$ where the atomic states are
rotated according to the rotation matrix (\ref{R}), that is,%
\begin{eqnarray}
\frac{1}{\sqrt{2}}( &\mid &f_{2}\rangle +\mid g_{2}\rangle )\rightarrow \mid
f_{2}\rangle ,  \nonumber \\
\frac{1}{\sqrt{2}}( &\mid &f_{2}\rangle -\mid g_{2}\rangle )\rightarrow
-\mid g_{2}\rangle ,
\end{eqnarray}%
and we get%
\begin{equation}
\mid \psi \rangle _{A1-A2-C}=\frac{1}{2}[\mid f_{1}\rangle \mid f_{2}\rangle
(|\alpha \rangle +|-\alpha \rangle )+\mid g_{1}\rangle \mid g_{2}\rangle
(|\alpha \rangle -|-\alpha \rangle )].  \label{A1A2C}
\end{equation}%
If we inject $|\alpha \rangle $ in cavity $C$ which mathematically is
represented by the operation $D(\beta )|\alpha \rangle =|\alpha +\beta
\rangle $ \cite{Louisell} and, taking $\beta =\alpha $, this gives us%
\begin{equation}
\mid \psi \rangle _{A1-A2-C}=\frac{1}{2}[\mid f_{1}\rangle \mid f_{2}\rangle
(|2\alpha \rangle +|0\rangle )+\mid g_{1}\rangle \mid g_{2}\rangle (|2\alpha
\rangle -|0\rangle )].
\end{equation}%
In order to disentangle the atomic states of the cavity field state we now
send a two-level atom $A3$ resonant with the cavity, with $|b_{3}\rangle $
and $|a_{3}\rangle $ being the lower and upper levels respectively, through $%
C$. If $A3$ is sent in the lower state $|b_{3}\rangle $, under the
Jaynes-Cummings dynamics \cite{JCM} (see (\ref{UJC}) with $\Delta =0$) we
know that the state $|b_{3}\rangle |0\rangle $ does not evolve, however, the
state $|b_{3}\rangle |2\alpha \rangle $ evolves to $|a_{3}\rangle |\chi
_{a}\rangle +|b_{3}\rangle |\chi _{b}\rangle $, where $|\chi _{b}\rangle
=\sum\limits_{n}C_{n}\cos (gt\sqrt{n})|n\rangle $ and $|\chi _{a}\rangle
=-i\sum\limits_{n}C_{n+1}\sin (gt\sqrt{n+1})|n\rangle $ and $C_{n}=e^{-\frac{%
1}{2}|2\alpha |^{2}}(2\alpha )^{n}/\sqrt{n!}$. Then we get%
\begin{equation}
\mid \psi \rangle _{A1-A2-C}=\frac{1}{2}[\mid f_{1}\rangle \mid f_{2}\rangle
(|a_{3}\rangle |\chi _{a}\rangle +|b_{3}\rangle |\chi _{b}\rangle
+|b_{3}\rangle |0\rangle )+\mid g_{1}\rangle \mid g_{2}\rangle
(|a_{3}\rangle |\chi _{a}\rangle +|b_{3}\rangle |\chi _{b}\rangle
-|b_{3}\rangle |0\rangle )],
\end{equation}%
and if we detect atom $A3$ in state $|a_{3}\rangle $ finally we get the EPR
(or Bell) state%
\begin{equation}
\mid \Phi ^{+}\rangle _{A1-A2}=\frac{1}{\sqrt{2}}(\mid f_{1}\rangle \mid
f_{2}\rangle +\mid g_{1}\rangle \mid g_{2}\rangle ),  \label{EPRPSI+}
\end{equation}%
which is an entangled state of atoms $A1$ and $A2$, which in principle may
be far apart from each other.

In the above disentanglement process we can choose a coherent field with a
photon-number distribution with a sharp peak at average photon number $%
\langle n\rangle =|\alpha |^{2}$ so that, to a good approximation, $|\chi
_{b}\rangle \cong C_{\overline{n}}\cos (\sqrt{\overline{n}}g\tau )|\overline{%
n}\rangle $ and $|\chi _{a}\rangle \cong C_{\overline{n}}\sin (\sqrt{%
\overline{n}}g\tau )|\overline{n}\rangle $, where $\overline{n}$ is the
integer nearest $\langle n\rangle $, and we could choose, for instance $\ 
\sqrt{\overline{n}}g\tau =\pi /2$, so that we would have $|\chi _{a}\rangle
\cong C_{\overline{n}}|\overline{n}\rangle $ and $|\chi _{b}\rangle \cong 0$%
. In this case, atom $A3$ \ would be detected in state $|a_{3}\rangle $ with
almost $100\%$ of probability. Therefore, proceeding this way, we can
guarantee that the atomic and field states will be disentangled successfully
as we would like.

Notice that starting from (\ref{A1A2C}) if we had injected $|-\alpha \rangle 
$ in the cavity and detected $|a_{3}\rangle $ we would get the EPR state%
\begin{equation}
\mid \Phi ^{-}\rangle _{A1-A2}=\frac{1}{\sqrt{2}}(\mid f_{1}\rangle \mid
f_{2}\rangle -\mid g_{1}\rangle \mid g_{2}\rangle ).  \label{EPRPSI-}
\end{equation}%
Now, if we apply an extra rotation on the states of atom $A2$ in (\ref%
{EPRPSI+}) in a Ramsey cavity $R5,$ according to the rotation matrix%
\begin{equation}
R=\mid g_{2}\rangle \langle f_{2}|-\mid f_{2}\rangle \langle g_{2}|,
\label{EPRR5}
\end{equation}%
we get%
\begin{equation}
\mid \Psi ^{-}\rangle _{A1-A2}=\frac{1}{\sqrt{2}}(\mid f_{1}\rangle \mid
g_{2}\rangle -\mid g_{1}\rangle \mid f_{2}\rangle ),  \label{EPRPHI-}
\end{equation}%
and applying (\ref{EPRR5}) on (\ref{EPRPSI-}) we get%
\begin{equation}
\mid \Psi ^{+}\rangle _{A1-A2}=\frac{1}{\sqrt{2}}(\mid f_{1}\rangle \mid
g_{2}\rangle +\mid g_{1}\rangle \mid f_{2}\rangle ).  \label{EPRPHI+}
\end{equation}

The states (\ref{EPRPSI+}), (\ref{EPRPSI-}), (\ref{EPRPHI-}) and (\ref%
{EPRPHI+}) form a Bell basis \cite{BELLbasis, Nielsen} which is a complete
orthonormal basis for atoms $A1$ and $A2$. These states show that quantum
entanglement implies non-locality. The manifestation of non-locality shows
up when we perform a measurement on one of the atoms. For instance, from (%
\ref{EPRPSI+}) it is clear that if we detect atom $A1$ in state $\mid
f_{1}\rangle $ then atom $A2$ collapses instantaneously to the state $\mid
f_{2}\rangle $ and if we detect atom $A1$ in state $\mid g_{1}\rangle $ then
atom $A2$ collapses instantaneously to the state $\mid g_{2}\rangle $, no
matter how distant they are from each other. The same applies to the other
states (\ref{EPRPSI-}), (\ref{EPRPHI-}) and (\ref{EPRPHI+}).

\section{THE GHZ EXPERIMENT\protect\bigskip}

In this section we are first going to show how to prepare an atomic GHZ
state of the Mermin kind \cite{Mermin} and then show in detail how to
perform the GHZ test once we have a GHZ state. Considering the states (\ref%
{EOCS}) we can write the state (\ref{A1A2C}) as%
\begin{equation}
\mid \psi ,+\rangle _{A1-A2-C}=\frac{1}{2}(\mid f_{1}\rangle \mid
f_{2}\rangle |+\rangle +\mid g_{1}\rangle \mid g_{2}\rangle |-\rangle ),
\label{GHZA1A2EOCS+}
\end{equation}%
which is a GHZ state involving atomic states of $A1$ and $A2$ and the cavity
field state. It is easy also to obtain the GHZ state%
\begin{equation}
\mid \psi ,-\rangle _{A1-A2-C}=\frac{1}{2}(\mid f_{1}\rangle \mid
f_{2}\rangle |+\rangle -\mid g_{1}\rangle \mid g_{2}\rangle |-\rangle ),
\label{GHZA1A2EOCS-}
\end{equation}

Let us prepare a two-level atom $A3$ in the Ramsey cavity $R5.$ If atom $A3$
is initially in the state $\mid g_{3}\rangle $, according to the rotation
matrix (\ref{R}) we get%
\begin{equation}
\mid \psi \rangle _{A3}=\frac{1}{\sqrt{2}}(\mid f_{3}\rangle +\mid
g_{3}\rangle ).
\end{equation}%
We take into account (\ref{U}) with $\varphi =\pi $. Starting from (\ref%
{GHZA1A2EOCS+}), after the atom has passed through the cavity we get%
\begin{equation}
\mid \psi \rangle _{A1-A2-A3-C}=\frac{1}{2\sqrt{2}}[\mid f_{1}\rangle \mid
f_{2}\rangle (\mid f_{3}\rangle +\mid g_{3}\rangle )(|\alpha \rangle
+|-\alpha \rangle )-\mid g_{1}\rangle \mid g_{2}\rangle (\mid f_{3}\rangle
-\mid g_{3}\rangle )(|\alpha \rangle -|-\alpha \rangle )].
\end{equation}%
Then, we let atom $A3$ to enter a Ramsey cavity $R6$ where the atomic states
are rotated according the rotation matrix (\ref{R}), which gives us 
\begin{eqnarray}
\frac{1}{\sqrt{2}}( &\mid &f_{3}\rangle +\mid g_{3}\rangle )\rightarrow \mid
f_{3}\rangle ,  \nonumber \\
\frac{1}{\sqrt{2}}( &\mid &f_{3}\rangle -\mid g_{3}\rangle )\rightarrow
-\mid g_{3}\rangle ,
\end{eqnarray}%
and we get%
\begin{equation}
\mid \psi \rangle _{A1-A2-A3-C}=\frac{1}{2}[\mid f_{1}\rangle \mid
f_{2}\rangle \mid f_{3}\rangle (|\alpha \rangle +|-\alpha \rangle )+\mid
g_{1}\rangle \mid g_{2}\rangle \mid g_{3}\rangle (|\alpha \rangle -|-\alpha
\rangle )].  \label{PSIA1A2A3C}
\end{equation}%
Injecting $|\alpha \rangle $ in cavity $\ C$ we have%
\begin{equation}
\mid \psi \rangle _{A1-A2-A3-C}=\frac{1}{2}[\mid f_{1}\rangle \mid
f_{2}\rangle \mid f_{3}\rangle (|2\alpha \rangle +|0\rangle )+\mid
g_{1}\rangle \mid g_{2}\rangle \mid g_{3}\rangle (|2\alpha \rangle
-|0\rangle )].
\end{equation}%
Now, we follow the same prescription used in the previous section in order
to disentangle the atomic states of the cavity field state. That is, we send
a two-level atom $A4$ resonant with the cavity through $C$ in the lower
state $|b_{4}\rangle $ and we detect atom $A4$ in the upper state $%
|a_{4}\rangle .$ Then finally we get the GHZ state%
\begin{equation}
\mid \psi ,+\rangle _{A1-A2-A3}=\frac{1}{\sqrt{2}}(\mid f_{1}\rangle \mid
f_{2}\rangle \mid f_{3}\rangle +\mid g_{1}\rangle \mid g_{2}\rangle \mid
g_{3}\rangle ).  \label{GHZ+}
\end{equation}

Notice that, if starting from (\ref{PSIA1A2A3C}), we had injected $|-\alpha
\rangle $ and detected $|a_{4}\rangle ,$ we would obtain the other GHZ state 
\begin{equation}
\mid \psi ,-\rangle _{A1-A2-A3}=\frac{1}{\sqrt{2}}(\mid f_{1}\rangle \mid
f_{2}\rangle \mid f_{3}\rangle -\mid g_{1}\rangle \mid g_{2}\rangle \mid
g_{3}\rangle ).  \label{GHZ-}
\end{equation}

Let us now first discuss a summary of the GHZ test prescription. We will
follow closely the discussion presented in the very clear article by Mermin
\ \cite{Mermin}. First we define the atomic operators%
\begin{eqnarray}
A &=&\sigma _{x}^{1}\sigma _{y}^{2}\sigma _{y}^{3},  \nonumber \\
B &=&\sigma _{y}^{1}\sigma _{x}^{2}\sigma _{y}^{3},  \nonumber \\
C &=&\sigma _{y}^{1}\sigma _{y}^{2}\sigma _{x}^{3},  \nonumber \\
D &=&\sigma _{x}^{1}\sigma _{x}^{2}\sigma _{x}^{3},  \label{ABCD}
\end{eqnarray}%
where%
\begin{eqnarray}
\sigma _{x}^{k} &=&\mid f_{k}\rangle \langle g_{k}\mid +\mid g_{k}\rangle
\langle f_{k}\mid ,  \nonumber \\
\sigma _{y}^{k} &=&-i(\mid f_{k}\rangle \langle g_{k}\mid -\mid g_{k}\rangle
\langle f_{k}\mid ),
\end{eqnarray}%
$(k=1,2$ and $3)$. It is easy to show that 
\begin{equation}
\lbrack A,b]=[A,C]=[B,C]=0,
\end{equation}%
and that%
\begin{equation}
A\mid \psi ,\pm \rangle _{A1-A2-A3}=B\mid \psi ,\pm \rangle
_{A1-A2-A3}=C\mid \psi ,\pm \rangle _{A1-A2-A3}=\mp 1\mid \psi ,\pm \rangle
_{A1-A2-A3},
\end{equation}%
and%
\begin{equation}
D\mid \psi ,\pm \rangle _{A1-A2-A3}=\pm 1\mid \psi ,\pm \rangle _{A1-A2-A3}.
\label{DPSI}
\end{equation}

If we assume that there are six elements of reality $m_{x}^{k}$ and $%
m_{y}^{k}$ $(k=1,2$ and $3)$ each having value $+1$ or $-1$ waiting to be
revealed according to a local realism theory, then we can write%
\begin{eqnarray}
a_{\pm } &=&m_{x}^{1}m_{y}^{2}m_{y}^{3}=\mp 1,  \nonumber \\
b_{\pm } &=&m_{y}^{1}m_{x}^{2}m_{y}^{3}=\mp 1,  \nonumber \\
c_{\pm } &=&m_{y}^{1}m_{y}^{2}m_{x}^{3}=\mp 1,
\end{eqnarray}%
and we have%
\begin{equation}
d_{\pm }=a_{\pm }b_{\pm }c_{\pm }=m_{x}^{1}m_{x}^{2}m_{x}^{3}=\mp 1,
\end{equation}%
where the upper and lower subindexes refer to the GHZ state (\ref{GHZ+}) and
(\ref{GHZ-}) respectively and we have used $(m_{y}^{k})^{2}=1$. So, the
existence of elements of reality implies that if we measure the value of the
observables $\sigma _{x}^{k}$ $(k=1,2$ and $3)$ (that is the elements of
reality associated with them) in the state $\mid \psi ,\pm \rangle
_{A1-A2-A3}$ the product of the three resulting values must be $d_{\pm }=\mp
1$. But according to (\ref{DPSI}) the eigenvalue of the operator $D$ applied
the state $\mid \psi ,\pm \rangle _{A1-A2-A3}$ is $\pm 1$. Therefore,
measuring this eigenvalue we can decide between theories based on local
realism and quantum mechanics.

Now, let us see how we proceed to perform the GHZ test (we follow closely
the scheme presented in \cite{GHZLambdaat}). We start letting atoms $Ak$ to
pass through the Ramsey zones $Kk$ \ ($k=1,2$ and $3)$ where the atomic
states are rotated according to the rotation matrix

\begin{equation}
K_{k}=\frac{1}{\sqrt{2}}\left[ 
\begin{array}{cc}
1 & -1 \\ 
1 & 1%
\end{array}%
\right] ,
\end{equation}%
or%
\begin{equation}
K_{k}=\frac{1}{\sqrt{2}}(\mid f_{k}\rangle \langle f_{k}\mid -\mid
f_{k}\rangle \langle g_{k}\mid +\mid g_{k}\rangle \langle f_{k}\mid +\mid
g_{k}\rangle \langle g_{k}\mid ).  \label{Kk}
\end{equation}%
These rotation matrixes are the key ingredient which allow us to perform the
GHZ test. The method of the test we are going to describe is based on a
gradual unraveling of the GHZ state being considered.

The eigenvalues of the operators $\sigma _{x}^{k}$ are%
\begin{equation}
|\psi _{x}^{k},\pm \rangle =\frac{1}{\sqrt{2}}(\mid f_{k}\rangle \pm \mid
g_{k}\rangle ).  \label{PSIx}
\end{equation}%
Let us take the state (\ref{GHZ+}). Writing the states $\mid f_{1}\rangle $
and $\mid g_{1}\rangle $ in terms of the states (\ref{PSIx}) for $k=1$ and
substituting in the GHZ state (\ref{GHZ+}) we get

\begin{equation}
\mid \psi ,+\rangle _{A1-A2-A3}=\frac{1}{2}[|\psi _{x}^{1},+\rangle (\mid
f_{2}\rangle \mid f_{3}\rangle +\mid g_{2}\rangle \mid g_{3}\rangle )+|\psi
_{x}^{1},-\rangle (\mid f_{2}\rangle \mid f_{3}\rangle -\mid g_{2}\rangle
\mid g_{3}\rangle )].  \label{PSIX1}
\end{equation}%
Applying (\ref{Kk}) to the state (\ref{GHZ+}) for $k=1$, we have%
\begin{equation}
K_{1}\mid \psi ,+\rangle _{A1-A2-A3}=\frac{1}{2}[|f_{1}\rangle (\mid
f_{2}\rangle \mid f_{3}\rangle -\mid g_{2}\rangle \mid g_{3}\rangle
)+|g_{1}\rangle (\mid f_{2}\rangle \mid f_{3}\rangle +\mid g_{2}\rangle \mid
g_{3}\rangle )].  \label{K1PSI123}
\end{equation}%
Now, we compare (\ref{K1PSI123}) and (\ref{PSIX1}). We see that the rotation
by $K_{1}$ followed by the detection of $|g_{1}\rangle $ corresponds to the
detection of the the state $|\psi _{x}^{1},+\rangle $ whose eigenvalue of $%
\sigma _{x}^{1}$ is $+1$. After we detect $|g_{1}\rangle $, we get%
\begin{equation}
\mid \psi \rangle _{A2-A3}=\frac{1}{\sqrt{2}}(\mid f_{2}\rangle \mid
f_{3}\rangle +\mid g_{2}\rangle \mid g_{3}\rangle ).  \label{PSI23}
\end{equation}%
As before, we rewrite the states $\mid f_{2}\rangle $ and $\mid g_{2}\rangle 
$ in terms of these states (\ref{PSIx}) for $k=2$ and substitute in the
above state and we have 
\begin{equation}
\mid \psi \rangle _{A2-A3}=\frac{1}{2}[|\psi _{x}^{2},+\rangle (\mid
f_{3}\rangle +\mid g_{3}\rangle )+|\psi _{x}^{2},-\rangle (\mid f_{3}\rangle
-\mid g_{3}\rangle )].  \label{PSIX2}
\end{equation}%
If we apply (\ref{Kk}) for $k=2$ to the state (\ref{PSI23}) we get%
\begin{equation}
K_{2}\mid \psi \rangle _{A2-A3}=\frac{1}{2}[|f_{2}\rangle (\mid f_{3}\rangle
-\mid g_{3}\rangle )+|g_{2}\rangle (\mid f_{3}\rangle +\mid g_{3}\rangle )].
\label{K2PSI23}
\end{equation}%
Again, if we compare (\ref{K2PSI23}) with (\ref{PSIX2}), we see that the
rotation by $K_{2}$ followed by the detection of $|g_{2}\rangle $
corresponds to the detection of the the state $|\psi _{x}^{2},+\rangle $
whose eigenvalue of $\sigma _{x}^{2}$ is $+1$. After we detect $%
|g_{2}\rangle $, we get%
\begin{equation}
\mid \psi \rangle _{A3}=\frac{1}{\sqrt{2}}(\mid f_{3}\rangle +\mid
g_{3}\rangle )=|\psi _{x}^{3},+\rangle ,  \label{PSIA3}
\end{equation}%
which is the eigenvector of $\sigma _{x}^{3}$ with eigenvalue $+1$. Now, we
finally apply (\ref{Kk}) for $k=3$ to the above state and we have%
\begin{equation}
K_{3}\mid \psi \rangle _{A3}=\mid g_{3}\rangle ,
\end{equation}%
and the only possibility is to detect atom $A3$ in the state $\mid
g_{3}\rangle .$

We can repeat the above procedure and see that we have only four
possibilities which are presented schematically below, where on the left, we
present the possible sequences of atomic state rotations through $K_{k}$ and
detections of $\mid f_{k}\rangle $ or $\mid g_{k}\rangle $ and on the right,
we present the sequences of the corresponding states $|\psi _{x}^{k},\pm
\rangle $ where $k=1,2$ and $3$ which corresponds to the measurement of the
eigenvalue of the operator $D$ given by (\ref{ABCD}), 
\begin{eqnarray}
(K_{1}, &\mid &g_{1}\rangle )(K_{2},\mid g_{2}\rangle )(K_{3},\mid
g_{3}\rangle )\longleftrightarrow |\psi _{x}^{1},+\rangle |\psi
_{x}^{2},+\rangle |\psi _{x}^{3},+\rangle ,  \nonumber \\
(K_{1}, &\mid &g_{1}\rangle )(K_{2},\mid f_{2}\rangle )(K_{3},\mid
f_{3}\rangle )\longleftrightarrow |\psi _{x}^{1},+\rangle |\psi
_{x}^{2},-\rangle |\psi _{x}^{3},-\rangle ,  \nonumber \\
(K_{1}, &\mid &f_{1}\rangle )(K_{2},\mid f_{2}\rangle )(K_{3},\mid
g_{3}\rangle )\longleftrightarrow |\psi _{x}^{1},-\rangle |\psi
_{x}^{2},-\rangle |\psi _{x}^{3},+\rangle ,  \nonumber \\
(K_{1}, &\mid &f_{1}\rangle )(K_{2},\mid g_{2}\rangle )(K_{3},\mid
f_{3}\rangle )\longleftrightarrow |\psi _{x}^{1},-\rangle |\psi
_{x}^{2},+\rangle |\psi _{x}^{3},-\rangle .  \label{GHZT+}
\end{eqnarray}%
Notice that all these results give us the eigenvalue $+1$ of the operator $D$
(see (\ref{DPSI})). Therefore, in one run, if we succeed to perform the
above rotations of the atomic states according to (\ref{Kk}), each one
followed by the detection of the respective state $\mid f_{k}\rangle $ or $%
\mid g_{k}\rangle $, we get the result of the experiment in favor of quantum
mechanics.

If we perform the test for the state (\ref{GHZ-}), following the above
procedure, it is easy to see that we have the four possible outcomes,%
\begin{eqnarray}
(K_{1}, &\mid &g_{1}\rangle )(K_{2},\mid g_{2}\rangle )(K_{3},\mid
f_{3}\rangle )\longleftrightarrow |\psi _{x}^{1},+\rangle |\psi
_{x}^{2},+\rangle |\psi _{x}^{3},-\rangle ,  \nonumber \\
(K_{1}, &\mid &g_{1}\rangle )(K_{2},\mid f_{2}\rangle )(K_{3},\mid
g_{3}\rangle )\longleftrightarrow |\psi _{x}^{1},+\rangle |\psi
_{x}^{2},-\rangle |\psi _{x}^{3},+\rangle ,  \nonumber \\
(K_{1}, &\mid &f_{1}\rangle )(K_{2},\mid f_{2}\rangle )(K_{3},\mid
f_{3}\rangle )\longleftrightarrow |\psi _{x}^{1},-\rangle |\psi
_{x}^{2},-\rangle |\psi _{x}^{3},-\rangle ,  \nonumber \\
(K_{1}, &\mid &f_{1}\rangle )(K_{2},\mid g_{2}\rangle )(K_{3},\mid
g_{3}\rangle )\longleftrightarrow |\psi _{x}^{1},-\rangle |\psi
_{x}^{2},+\rangle |\psi _{x}^{3},+\rangle .  \label{GHZT-}
\end{eqnarray}%
Notice that all these results give us the eigenvalue $-1$ of the operator $D$
(see (\ref{DPSI})), as it should be. See Fig. 2 where we present a scheme of
the GHZ experiment.

Notice that if we start from (\ref{GHZA1A2EOCS+}), following the
prescription (\ref{GHZT+}) and (\ref{GHZT-}), we do not have to apply $K_{3}$
and detect $\mid f_{3}\rangle $ or $\mid g_{3}\rangle $ but instead, after
applying $K_{2}$ \ followed by the corresponding detection of the state of $%
A2,$ we would get the cavity in the state $|\psi _{x}^{3},+\rangle =$ $\mid
\alpha \rangle $ or $|\psi _{x}^{3},-\rangle =$ $\mid -\alpha \rangle $.
Putting in another way, if we define%
\begin{equation}
D=\sigma _{x}^{1}\sigma _{x}^{2}\sigma _{x}^{C},
\end{equation}%
where%
\begin{equation}
\sigma _{x}^{C}=|+\rangle \langle -|+|-\rangle \langle +|  \label{SigmaxC}
\end{equation}%
and $|+\rangle $ and $|-\rangle $ are given by (\ref{EOCS}), we have for the
eigenvectors of $\sigma _{x}^{C}$ 
\begin{eqnarray}
|\psi _{x}^{3},+\rangle &=&|\psi _{x}^{C},+\rangle =\frac{1}{2}(|+\rangle
+|-\rangle )=\mid \alpha \rangle  \nonumber \\
|\psi _{x}^{3},-\rangle &=&|\psi _{x}^{C},-\rangle =\frac{1}{2}(|+\rangle
-|-\rangle )=\mid -\alpha \rangle  \label{EVsigmaxC}
\end{eqnarray}%
Then, in order to conclude the GHZ test, in the case in which the cavity is
left in the state $|\psi _{x}^{C},+\rangle =$ $\mid \alpha \rangle $ we
could send an atom $A3$ in the state 
\begin{equation}
\mid \psi \rangle _{A3}=\frac{1}{\sqrt{2}}(\mid f_{3}\rangle +\mid
g_{3}\rangle ),
\end{equation}%
through $C$. After atom $A3$ interacts with the cavity we would have%
\begin{equation}
\mid \psi \rangle _{A3-C}=\frac{1}{\sqrt{2}}(\mid f_{3}\rangle \mid -\alpha
\rangle +\mid g_{3}\rangle \mid \alpha \rangle ),
\end{equation}%
Now, we follow the same prescription used in the previous section in order
to disentangle the atomic states of the cavity field state. That is, we
inject $\mid \alpha \rangle $ in the cavity and we get 
\begin{equation}
\mid \psi \rangle _{A3-C}=\frac{1}{\sqrt{2}}(\mid f_{3}\rangle \mid 0\rangle
+\mid g_{3}\rangle \mid 2a\rangle ),
\end{equation}%
and we send a two-level atom $A4$ resonant with the cavity through $C$ in
the lower state $|b_{4}\rangle $ and if we detect atom $A4$ in the upper
state $|a_{4}\rangle $ we have 
\begin{equation}
\mid \psi \rangle _{A3-C}=\mid g_{3}\rangle |\chi _{a}\rangle ,
\end{equation}%
and the detection of $\mid g_{3}\rangle $ is the signature that the cavity
was previously in the state $\mid \alpha \rangle $. If the cavity was in the
state $\mid -\alpha \rangle $ after \ atom $A3$ has interacted with the
cavity we have%
\begin{equation}
\mid \psi \rangle _{A3-C}=\frac{1}{\sqrt{2}}(\mid f_{3}\rangle \mid \alpha
\rangle +\mid g_{3}\rangle \mid -\alpha \rangle ).
\end{equation}%
Now, we inject $\mid \alpha \rangle $ and we have%
\begin{equation}
\mid \psi \rangle _{A3-C}=\frac{1}{\sqrt{2}}(\mid f_{3}\rangle \mid 2\alpha
\rangle +\mid g_{3}\rangle \mid 0\rangle ).
\end{equation}%
and we send a two-level atom $A4$ resonant with the cavity through $C$ in
the lower state $|b_{4}\rangle $ and if we detect atom $A4$ in the upper
state $|a_{4}\rangle $ we have 
\begin{equation}
\mid \psi \rangle _{A3-C}=\mid f_{3}\rangle |\chi _{a}\rangle ,
\end{equation}%
and the detection of $\mid f_{3}\rangle $ is the signature that the cavity
was previously in the state $\mid -\alpha \rangle $. \ With the detection of 
$|a_{4}\rangle $ \ and \ $\mid g_{3}\rangle $ or $|a_{4}\rangle $ \ and \ $%
\mid f_{3}\rangle $ \ we conclude the GHZ test based on the GHZ state (\ref%
{GHZA1A2EOCS+}) or state (\ref{GHZA1A2EOCS-}).

We summarize below the results of the GHZ test: if we perform this test for
the state (\ref{GHZA1A2EOCS+}), following the above procedure, it is easy to
see that we have the four possible outcomes, 
\begin{eqnarray}
(K_{1}, &\mid &g_{1}\rangle )(K_{2},\mid g_{2}\rangle )(|a_{4}\rangle \
)(\mid g_{3}\rangle )\longleftrightarrow |\psi _{x}^{1},+\rangle |\psi
_{x}^{2},+\rangle |\psi _{x}^{3},+\rangle ,  \nonumber \\
(K_{1}, &\mid &g_{1}\rangle )(K_{2},\mid f_{2}\rangle )(|a_{4}\rangle \
)(\mid f_{3}\rangle )\longleftrightarrow |\psi _{x}^{1},+\rangle |\psi
_{x}^{2},-\rangle |\psi _{x}^{3},-\rangle ,  \nonumber \\
(K_{1}, &\mid &f_{1}\rangle )(K_{2},\mid f_{2}\rangle )(|a_{4}\rangle \
)(\mid g_{3}\rangle )\longleftrightarrow |\psi _{x}^{1},-\rangle |\psi
_{x}^{2},-\rangle |\psi _{x}^{3},+\rangle ,  \nonumber \\
(K_{1}, &\mid &f_{1}\rangle )(K_{2},\mid g_{2}\rangle )(|a_{4}\rangle \
)(\mid f_{3}\rangle )\longleftrightarrow |\psi _{x}^{1},-\rangle |\psi
_{x}^{2},+\rangle |\psi _{x}^{3},-\rangle .
\end{eqnarray}%
Notice that all these results give us the eigenvalue $+1$ of the operator $D$
(see (\ref{DPSI})).

If we perform the test for the state (\ref{GHZA1A2EOCS-}), following the
above procedure, it is easy to see that we have the four possible outcomes,%
\begin{eqnarray}
(K_{1}, &\mid &g_{1}\rangle )(K_{2},\mid g_{2}\rangle )(|a_{4}\rangle \
)(\mid f_{3}\rangle )\longleftrightarrow |\psi _{x}^{1},+\rangle |\psi
_{x}^{2},+\rangle |\psi _{x}^{3},-\rangle ,  \nonumber \\
(K_{1}, &\mid &g_{1}\rangle )(K_{2},\mid f_{2}\rangle )(|a_{4}\rangle \
)(\mid g_{3}\rangle )\longleftrightarrow |\psi _{x}^{1},+\rangle |\psi
_{x}^{2},-\rangle |\psi _{x}^{3},+\rangle ,  \nonumber \\
(K_{1}, &\mid &f_{1}\rangle )(K_{2},\mid f_{2}\rangle )(|a_{4}\rangle \
)(\mid f_{3}\rangle )\longleftrightarrow |\psi _{x}^{1},-\rangle |\psi
_{x}^{2},-\rangle |\psi _{x}^{3},-\rangle ,  \nonumber \\
(K_{1}, &\mid &f_{1}\rangle )(K_{2},\mid g_{2}\rangle )(|a_{4}\rangle \
)(\mid g_{3}\rangle )\longleftrightarrow |\psi _{x}^{1},-\rangle |\psi
_{x}^{2},+\rangle |\psi _{x}^{3},+\rangle .
\end{eqnarray}%
Notice that all these results give us the eigenvalue $-1$ of the operator $D$
(see (\ref{DPSI})), as it should be. See Fig. 3 where we present a scheme of
this GHZ experiment.

\section{CONCLUSION}

Concluding, we have presented a scheme of realization of atomic GHZ state
and the GHZ test making use of cavity QED.\ In the scheme presented here we
use atoms interacting with a superconducting cavity prepared in a coherent
state which is a state relatively easy to be prepared and handled. In our
scheme we make use of atoms in a cascade configuration and in the scheme
presented in \cite{GHZLambdaat} we make use of atoms in a lambda
configuration. The advantage of using a cascade atomic configuration is that
the atomic state rotation and detection process is simpler than in the
lambda configuration where we have states which are degenerated. On the
other hand, for the cascade configuration we have to perform more rotations
of the atomic states using Ramsey cavities than in the case of the lambda
configuration.

\appendix

\section{\protect\bigskip Time evolution operator for two-level atoms}

Let us consider a two-level atom interacting with a cavity field, where $%
|e\rangle $ and $|f\rangle $ are the upper and lower states respectively,
with $\omega _{e}$ and $\omega _{f}$ being the two atomic frequencies
associated to these two states and $\omega $ the cavity field frequency (see
Fig. 1). The Jaynes-Cummings Hamiltonian, under the rotating-wave
approximation, is given by \cite{JCM, Orszag, WallsMilburn}%
\begin{equation}
H=\hbar a^{\dag }a+\hbar \omega _{e}|e\rangle \langle e|++\hbar \omega
_{f}|f\rangle \langle f|+\hbar g(a|e\rangle \langle f|+a^{\dag }|f\rangle
\langle e|),  \label{JCH}
\end{equation}%
where $a^{\dag }$ and $a$ are the creation and annihilation operators
respectively for the cavity field, $g$ is the coupling constant and we write%
\begin{equation}
H=H_{0}+H_{I},
\end{equation}%
where we have settled%
\begin{eqnarray}
H_{0} &=&\hbar a^{\dag }a+\hbar \omega _{e}|e\rangle \langle e|++\hbar
\omega _{f}|f\rangle \langle f|,  \nonumber \\
H_{I} &=&\hbar g(a|e\rangle \langle f|+a^{\dag }|f\rangle \langle e|).
\end{eqnarray}%
Let us define the interaction picture%
\begin{equation}
|\psi _{I}\rangle =e^{i\frac{H_{0}}{\hbar }t}|\psi _{S}\rangle .
\end{equation}%
Taking into account%
\begin{equation}
i\hbar \frac{d}{dt}|\psi _{S}\rangle =H|\psi _{S}\rangle ,
\end{equation}%
we get%
\begin{equation}
i\hbar \frac{d}{dt}|\psi _{I}\rangle =V_{I}|\psi _{I}\rangle ,
\end{equation}%
where%
\begin{equation}
V_{I}=e^{i\frac{H_{0}}{\hbar }t}H_{I}e^{-i\frac{H_{0}}{\hbar }t}=\hbar \left[
\begin{array}{cc}
0 & ge^{i\Delta t}a \\ 
ge^{-i\Delta t}a^{\dagger } & 0%
\end{array}%
\right] ,
\end{equation}%
and%
\begin{equation}
\Delta =(\omega _{e}-\omega _{f})-\omega .
\end{equation}%
Considering%
\begin{equation}
|\psi _{I}(t)\rangle =U_{I}(t)|\psi _{I}(0)\rangle =U_{I}(t)|\psi
_{S}(0)\rangle ,
\end{equation}%
we have to solve the Schr\"{o}dinger's equation for the time evolution
operator%
\begin{equation}
i\hbar \frac{dU_{I}}{dt}=V_{I}U_{I},  \label{AP2}
\end{equation}%
where%
\begin{equation}
U_{I}(t)=\left[ 
\begin{array}{cc}
u_{ee}(t) & u_{ef}(t) \\ 
u_{fe}(t) & u_{ff}(t)%
\end{array}%
\right]
\end{equation}%
and%
\begin{equation}
U_{I}(0)=\left[ 
\begin{array}{cc}
1 & 0 \\ 
0 & 1%
\end{array}%
\right] .
\end{equation}%
That is,%
\begin{eqnarray}
i\frac{d}{dt}u_{ee}(t) &=&ge^{i\Delta t}au_{ef}(t),  \nonumber \\
i\frac{d}{dt}u_{ef}(t) &=&ge^{i\Delta t}au_{ff}(t),  \nonumber \\
i\frac{d}{dt}u_{fe}(t) &=&ge^{-i\Delta t}a^{\dagger }u_{ee}(t),  \nonumber \\
i\frac{d}{dt}u_{ff}(t) &=&ge^{-i\Delta t}a^{\dagger }u_{ef}(t).  \label{EDI4}
\end{eqnarray}%
Now, we are going to use the abbreviation%
\begin{equation}
\alpha =i\frac{\Delta }{2}
\end{equation}%
and we have%
\begin{eqnarray}
ie^{-\alpha t}\frac{d}{dt}u_{ee}(t) &=&ge^{\alpha t}au_{ef}(t),  \nonumber \\
ie^{-\alpha t}\frac{d}{dt}u_{ef}(t) &=&ge^{\alpha t}au_{ff}(t),  \nonumber \\
ie^{\alpha t}\frac{d}{dt}u_{fe}(t) &=&ge^{-\alpha t}a^{\dagger }u_{ee}(t), 
\nonumber \\
ie^{\alpha t}\frac{d}{dt}u_{ff}(t) &=&ge^{-\alpha t}a^{\dagger }u_{ef}(t),
\end{eqnarray}%
which can be solved easily using, for instance, Laplace transformation ($%
L\{u(t)\}=\widetilde{u}(s)$) and we have 
\begin{eqnarray}
i[(s+\alpha )\widetilde{u_{ee}}(s+\alpha )-1] &=&ga\widetilde{u_{ef}}%
(s-\alpha ),  \nonumber \\
i(s+\alpha )\widetilde{u_{ef}}(s+\alpha ) &=&ga\widetilde{u_{ff}}(s-\alpha ),
\nonumber \\
i(s-\alpha )\widetilde{u_{fe}}(s-\alpha ) &=&ga^{\dagger }\widetilde{u_{ee}}%
(s+\alpha ),  \nonumber \\
i[(s-\alpha )\widetilde{u_{ff}}(s-\alpha )-1] &=&ga^{\dagger }\widetilde{%
u_{ef}}(s+\alpha ),
\end{eqnarray}%
and solving these algebraic equations and taking the inverse Laplace
transformation we get%
\begin{equation}
U_{I}(t)=\left[ 
\begin{array}{cc}
e^{i\frac{\Delta }{2}t}(\cos \mu t-i\frac{\Delta }{2\mu }\sin \mu t) & 
-ige^{i\frac{\Delta }{2}t}\frac{1}{\mu }(\sin \mu t)a \\ 
-iga^{\dag }e^{-i\frac{\Delta }{2}t}\frac{1}{\mu }(\sin \mu t) & e^{-i\frac{%
\Delta }{2}t}(\cos \nu t+i\frac{\Delta }{2\nu }\sin \nu t)%
\end{array}%
\right] ,  \label{UJC}
\end{equation}%
where we have defined%
\begin{eqnarray}
\mu &=&\sqrt{g^{2}aa^{\dag }+\frac{\Delta ^{2}}{4}},  \nonumber \\
\nu &=&\sqrt{g^{2}a^{\dag }a+\frac{\Delta ^{2}}{4}}.  \label{AP4}
\end{eqnarray}

In the large detuning limit ($\Delta \gg g$) we have%
\begin{eqnarray}
\mu &=&\sqrt{g^{2}aa^{\dag }+\frac{\Delta ^{2}}{4}}\cong \frac{\Delta }{2}+%
\frac{g^{2}aa^{\dag }}{\Delta },  \nonumber \\
\nu &=&\sqrt{g^{2}a^{\dag }a+\frac{\Delta ^{2}}{4}}.\cong \frac{\Delta }{2}+%
\frac{g^{2}a^{\dag }a}{\Delta }.
\end{eqnarray}%
and we get easily 
\begin{equation}
U_{d}(t)=e^{-i\varphi (a^{\dagger }a+1)}\mid e\rangle \langle e\mid
+e^{i\varphi a^{\dagger }a}\mid f\rangle \langle f\mid ,  \label{AP5}
\end{equation}%
where $\varphi =$ $g^{2}t/\Delta .$ If we neglect spontaneous emition, the
time evolution operator can also be obtained from the effective Hamiltonian%
\begin{equation}
H_{d}=\hbar \frac{g^{2}}{\Delta }a^{\dagger }a(\mid e\rangle \langle e\mid
-\mid f\rangle \langle f\mid ),  \label{AP6}
\end{equation}%
which agrees with the expression presented in \cite{Gerry, Holland, Milburn,
WallsMilburn, Orszag}. The subindexes in (\ref{AP5}) and (\ref{AP6}) are
related to the atom-field interaction described by them, that is, a
dispersive interaction.\bigskip

\noindent \textbf{Figure Captions} \bigskip

\textbf{Fig. 1-} Energy states scheme of a three-level atom where $|e\rangle 
$ is the upper state with atomic frequency $\omega _{e}$, $\ |f\rangle $ is
the intermediate state with atomic frequency $\omega _{f}$, $|g\rangle $ is
the lower state with atomic frequency $\omega _{g}$ and $\omega $ is the
cavity field frequency and $\Delta =(\omega _{e}-\omega _{f})-\omega $ is
the detuning. The transition $\mid f\rangle \rightleftharpoons \mid e\rangle 
$ is far enough of resonance with the cavity central frequency such that
only virtual transitions occur between these levels (only these states
interact with field in cavity $C$). In addition we assume that the
transition $\mid e\rangle \rightleftharpoons \mid g\rangle $ is highly
detuned from the cavity frequency so that there will be no coupling with the
cavity field in $C$.\bigskip

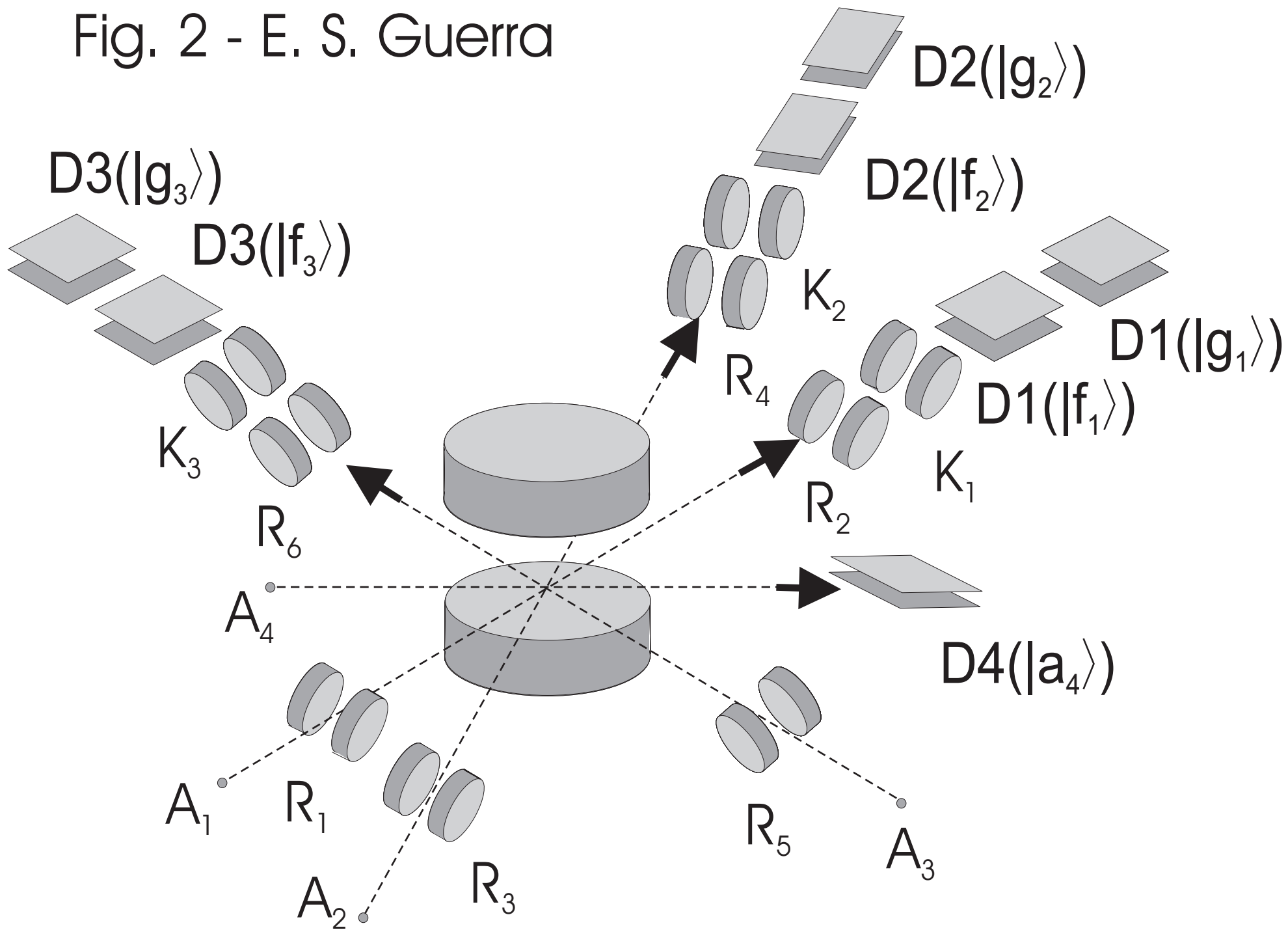
\textbf{Fig. 2}- Set-up for the GHZ experiment. Atom $A1$ passes through \
the Ramsey cavity $R1$ where it is prepared in a coherent superposition,
cavity $C$ and through the Ramsey cavity $R2.$ Atom $A2$ passes through \
the Ramsey cavity $R3$ where it is prepared in a coherent superposition,
cavity $C$ and through the Ramsey cavity $R4$. Atom $A3$ passes through \
the Ramsey cavity $R5$ where it is prepared in a coherent superposition, and
cavity $C$ through the Ramsey cavity $R6$. Then we inject a coherent state $%
|\alpha \rangle $ or $|-\alpha \rangle $ in the cavity and send a two-level
atom $A4$, initially in the lower state $|b_{4}\rangle $, resonant with
cavity $C$ through $C$ and then we detect $A4$ in detector $D4$ in the upper
state $|a_{4}\rangle $. Once the GHZ state has been obtained, the GHZ test
is performed making use of the Ramsey cavities $K1$, $K2$ and $K3$ and
detectors $D1$, $D2$ and $D3$ as described in the text.\bigskip

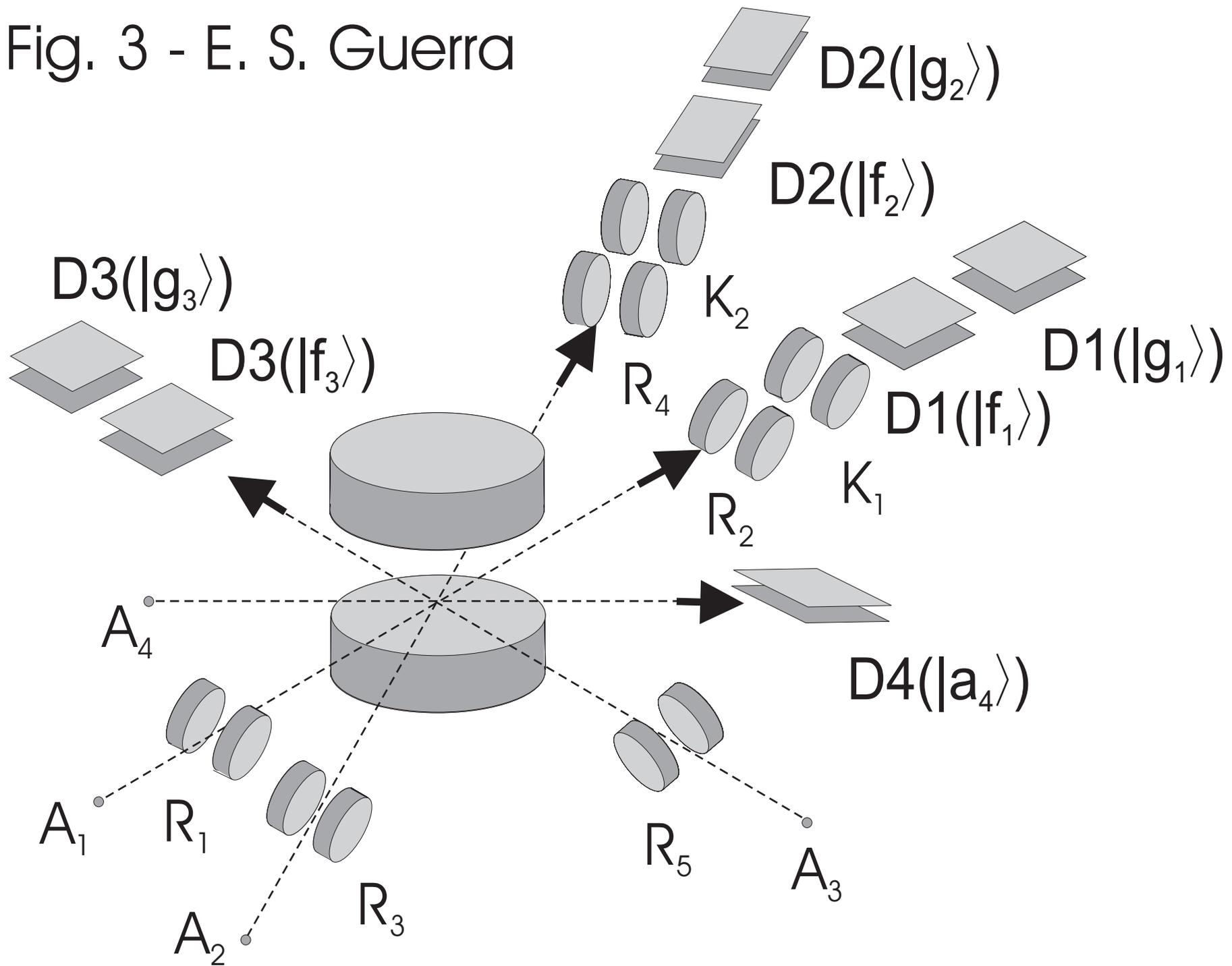
\textbf{Fig. 3}- Set-up for the GHZ experiment. Atom $A1$ passes through \
the Ramsey cavity $R1$ where it is prepared in a coherent superposition,
cavity $C$ and through the Ramsey cavity $R2.$ Atom $A2$ passes through \
the Ramsey cavity $R3$ where it is prepared in a coherent superposition,
cavity $C$ and through the Ramsey cavity $R4$. Atom $A3$ passes through \
the Ramsey cavity $R5$ where it is prepared in a coherent superposition, and
cavity $C$. Then we inject a coherent state $|\alpha \rangle $ or $|-\alpha
\rangle $ in the cavity and send a two-level atom $A4$, initially in the
lower state $|b_{4}\rangle $, resonant with cavity $C$ through $C$ and then
we detect $A4$ in detector $D4$ in the upper state $|a_{4}\rangle $. Once
the GHZ state has been obtained, the GHZ test is performed making use of the
Ramsey cavities $K1$ and $K2$ and detectors $D1$, $D2$ and $D3$ as described
in the text.

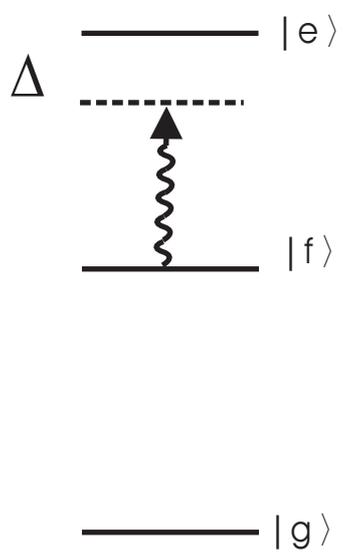
\end{document}